\documentclass[a4paper]{jpconf}
\usepackage{graphicx}
\usepackage{amsmath,amssymb,amsthm,amsxtra,overpic,bbm,bm,epsfig}

\begin{document}
\title{Majorana phases in neutrino-antineutrino oscillations}

\author{Ye-Ling Zhou}

\address{Institute of High Energy Physics, Chinese Academy of
Sciences, P.O. Box 918, Beijing 100049, China}

\ead{zhouyeling@ihep.ac.cn}

\begin{abstract}
If the massive neutrinos are Majorana particles, neutrinoless double beta
($0\nu\beta\beta$) decay experiments are not enough to determine the Majorana 
phases. We carry out a systematic study of CP violation in
neutrino-antineutrino oscillations. In these processes, CP-conserving parts
involve six independent $0\nu\beta\beta$-like mass terms $\langle
m\rangle^{}_{\alpha\beta}$ and CP-violating parts are associated
with nine independent Jarlskog-like parameters ${\cal
V}^{ij}_{\alpha\beta}$ (for $\alpha, \beta = e, \mu, \tau$ and $i, j
= 1, 2 ,3$). With the help of current neutrino oscillation data, we
illustrate the salient features of six independent
CP-violating asymmetries between $\nu^{}_\alpha \to
\overline{\nu}^{}_\beta$ and $\overline{\nu}^{}_\alpha \to
\nu^{}_\beta$ oscillations.
\end{abstract}

\section{Introduction}
The lepton flavor mixing is described by the Pontecorvo-Maki-Nakagawa-Sakata (PMNS) matrix $U$ \cite{PMNS}.Given neutrinos the Majorana nature, the PMNS matrix can
be parametrized in terms of three flavor mixing angles
($\theta^{}_{12}, \theta^{}_{13}, \theta^{}_{23}$) and three
CP-violating phases ($\delta, \rho, \sigma$) as follows:
\begin{eqnarray}
U \hspace{-0.2cm} & = & \hspace{-0.2cm} \left( \begin{matrix}
c^{}_{12} c^{}_{13} & s^{}_{12} c^{}_{13} & s^{}_{13} e^{-{\rm i}
\delta} \cr -s^{}_{12} c^{}_{23} - c^{}_{12} s^{}_{13} s^{}_{23}
e^{{\rm i} \delta} & c^{}_{12} c^{}_{23} - s^{}_{12} s^{}_{13}
s^{}_{23} e^{{\rm i} \delta} & c^{}_{13} s^{}_{23} \cr s^{}_{12}
s^{}_{23} - c^{}_{12} s^{}_{13} c^{}_{23} e^{{\rm i} \delta} &
-c^{}_{12} s^{}_{23} - s^{}_{12} s^{}_{13} c^{}_{23} e^{{\rm i}
\delta} & c^{}_{13} c^{}_{23} \cr
\end{matrix} \right) \left(
\begin{matrix} e^{{\rm i} \rho} & 0 & 0 \cr 0 & e^{{\rm i} \sigma}
& 0 \cr 0 & 0 & 1 \cr \end{matrix} \right) \; ,
%     (1)
\end{eqnarray}
where $c^{}_{ij} \equiv \cos\theta^{}_{ij}$ and $s^{}_{ij} \equiv
\sin\theta^{}_{ij}$ (for $ij = 12, 13, 23$). So far all the three neutrino mixing angles have been
measured to precisely in neutrino oscillation
experiments \cite{PDG}. A determination of the phase parameter
$\delta$ will be one of the major goals in the
next-generation long-baseline neutrino oscillation experiments. The
most challenging task is to detect the Majorana phases $\rho$ and
$\sigma$, which can only emerge in the lepton flavor violation (LNV) processes. Ref. \cite{Xing13} shows that it is in principle possible to determine all the three phases from the CP-violating
asymmetries ${\cal A}^{}_{\alpha\beta}$ between $\nu^{}_\alpha \to
\overline{\nu}^{}_\beta$ and $\overline{\nu}^{}_\alpha \to
\nu^{}_\beta$ oscillations. Here we report our work in Ref. \cite{XZ13}, where we give a systematic analysis of the effective parameters and CP-violating asymmetries in the neutrino-antineutrino oscillations.

In section 2 we briefly review some issues of three-flavor
neutrino-antineutrino oscillations. Six independent $0\nu\beta\beta$-like mass
terms $\langle m\rangle^{}_{\alpha\beta}$ and nine independent
Jalskog-like parameters ${\cal V}^{ij}_{\alpha\beta}$ (for $\alpha,
\beta = e, \mu, \tau$ and $i, j = 1, 2 ,3$) are introduced and their main features are discussed. In section 3 we study
the sensitivities of six CP-violating asymmetries ${\cal
A}^{}_{\alpha\beta}$ to the three phases, neutrino mass ordering, and the ratio of the
neutrino beam energy $E$ to the baseline length $L$. 
Section 4 is devoted to a summary of this report.

\section{Neutrino-antineutrino oscillations and effective parameters}

The neutrino-antineutrino oscillation probabilities for $\nu^{}_\alpha \to
\overline{\nu}^{}_\beta$ and $\overline{\nu}^{}_\alpha \to
\nu^{}_\beta$ are given by \cite{Xing13}
\vspace{-0.4cm}
\begin{eqnarray}
P(\nu^{}_\alpha \to \overline{\nu}^{}_\beta) \hspace{-0.2cm} & = &
\hspace{-0.2cm} \frac{|K|^2}{E^2} \Big[ \left| \langle m
\rangle^{}_{\alpha \beta} \right|^2 - 4 \sum_{i<j} m^{}_i m^{}_j
{\cal C}^{ij}_{\alpha\beta} \sin^2 \phi^{}_{ji} + 2 \sum_{i<j}
m^{}_i m^{}_j {\cal V}^{ij}_{\alpha\beta} \sin 2\phi^{}_{ji} \Big]
\; ,\nonumber \end{eqnarray}
\vspace{-0.9cm}
\begin{eqnarray}
P(\overline{\nu}^{}_\alpha \to \nu^{}_\beta) \hspace{-0.2cm} & = &
\hspace{-0.2cm} \frac{|\overline{K}|^2}{E^2} \Big[ \left| \langle
m \rangle^{}_{\alpha \beta} \right|^2 - 4 \sum_{i<j} m^{}_i m^{}_j
{\cal C}^{ij}_{\alpha\beta} \sin^2 \phi^{}_{ji} - 2 \sum_{i<j}
m^{}_i m^{}_j {\cal V}^{ij}_{\alpha\beta} \sin 2\phi^{}_{ji} \Big]
\; ,
%     (3)
\end{eqnarray}
\vspace{-0.5cm}\\
in which $\phi^{}_{ji} \equiv \Delta m^2_{ji} L/(4 E)$ with $\Delta
m^2_{ji} \equiv m^2_j - m^2_i$ being the mass-squared difference and $E$ the neutrino
(or antineutrino) beam energy, $L$ the baseline length.
$K$ and $\overline{K}$ are the kinematical factors, which satisfy $|K| = |\overline{K}|$. $m^{}_i/E$ stands for the helicity
suppression in the transition between $\nu^{}_i$ and
$\overline{\nu}^{}_i$. 
The effective mass term $\langle m
\rangle^{}_{\alpha \beta}$ is defined as 
$\langle m \rangle^{}_{\alpha \beta} \equiv \sum_i m^{}_i U^{}_{\alpha i} U^{}_{\beta i} $.
We call them $0\nu\beta\beta$-like
mass terms since $\langle m \rangle^{}_{ee}$ is the
effective mass term of the $0\nu\beta\beta$ decay. 
${\cal C}^{ij}_{\alpha\beta} \equiv  {\rm Re}(U^{}_{\alpha i} U^{}_{\beta i}
U^*_{\alpha j} U^*_{\beta j} ) $ and
${\cal V}^{ij}_{\alpha\beta} \equiv {\rm Im}(U^{}_{\alpha i} U^{}_{\beta i}
U^*_{\alpha j} U^*_{\beta j} ) $
describe the CP-conserving and CP-violating contributions of the PMNS
mixing, where the Greek and Latin subscripts running over $(e, \mu, \tau)$
and $(1, 2, 3)$, respectively. The CP-violating parts are referred to as Jarlskog-like parameters since they are similar to the definition of the Jarlskog parameter
${\cal J}  = {\rm Im} (U^{}_{e 1} U^{}_{\mu 2} U^*_{e 2}
U^*_{\mu 1} ) $ \cite{J}.

A measurement of neutrino-antineutrino
oscillations is far beyond nowadays experimental
capability as $P(\overline{\nu}^{}_\alpha \to
\nu^{}_\beta)$ highly suppressed by the factor $m^2_i/E^2$. The typical oscillation lengths by taking $E\sim {\cal O} \left(10\right)$ keV for example,
(1) $L^{\text{osc}}_{31} \simeq L^{\text{osc}}_{32} \simeq \frac{E}{10 ~\text{keV}}
\times 10 ~\text{m} $,
and (2) $L^{\text{osc}}_{21} \simeq \frac{E}{10 ~\text{keV}} \times 330 ~\text{m} $,
corresponding to $\Delta m^2_{21} \simeq 7.5 \times 10^{-5} ~{\rm
eV}^2$ and $|\Delta m^2_{31}| \simeq |\Delta m^2_{32}| \simeq 2.4
\times 10^{-3} ~{\rm eV}^2$, respectively. To make the oscillation effect detectable,
the sizes of the neutrino (or antineutrino) source and the detector
should be much smaller than $L^{\text{osc}}_{21}$ and (or) $L^{\text{osc}}_{31} \simeq
L^{\text{osc}}_{32}$. 

The Jarlskog-like parameters ${\cal
V}^{ij}_{\alpha\beta}$ satisfy the relations
${\cal V}^{ij}_{\alpha\beta}$=
${\cal V}^{ij}_{\beta\alpha}$ = $-{\cal V}^{ji}_{\alpha\beta}$ = $-{\cal
V}^{ji}_{\beta\alpha}$,
and ${\cal V}^{ii}_{\alpha\beta}$=$0$; but ${\cal
V}^{ij}_{\alpha\alpha} \neq 0$ for $i\neq j$. With the sum rules
${\cal V}^{ij}_{\alpha\beta} =({\cal V}^{ij}_{\gamma\gamma} - {\cal V}^{ij}_{\alpha\alpha} -
{\cal V}^{ij}_{\beta\beta})/2 $ ($\alpha\neq\beta\neq\gamma\neq\alpha$), we have only nine independent ${\cal V}^{ij}_{\alpha\beta}$. For a numerical illustration, we take $\theta^{}_{12} \simeq 33.4^\circ$, $\theta^{}_{13} \simeq
8.66^\circ$ and $\theta^{}_{23} \simeq 40.0^\circ$ as inputs \cite{FIT}. 
In general, ${\cal V}^{12}_{ee}$, ${\cal
V}^{23}_{\mu\mu}$, ${\cal V}^{23}_{\tau\tau}$ and ${\cal
V}^{23}_{\mu\tau}$ can maximally reach about $20\%$ in magnitude; in
comparison, the Jarlskog invariant ${\cal J} \leq 9.6\%$, suppressed by $s_{13}$. 
One special case is the ``pseudo-Dirac" case with $\rho = \sigma = 0$. 
This case is interesting because appreciable CP- and T-violating
effects are expected to show up in neutrino-antineutrino
oscillations even if the Majorana phases $\rho$ and $\sigma$ vanish.

The effective mass terms $\langle m\rangle^{}_{\alpha\beta}$ are important to understand the origin of neutrino
masses, since they are simply the $(\alpha, \beta)$ elements of the
symmetric Majorana neutrino mass matrix $M^{}_\nu$ in the charged lepton flavor basis.
These parameters also play important roles in some other LNV processes, such as the doubly charged Higgs decay $H^{++} \to \ell^+_\alpha \ell^+_\beta$ \cite{Han} in the type-II seesaw mechanism \cite{SS2} and some LNV decays of $K$, $D$ and $B$ mesons \cite{Valle}.
A measurement of the three CP-violating phases is
absolutely necessary in order to fully reconstruct the neutrino mass
matrix $M^{}_\nu$. Figure 1 illustrates the profiles of six $|\langle
m\rangle^{}_{\alpha\beta}|$, with inputs shown above. We allow $\delta$ to randomly vary in $(0,360^\circ)$ and $\rho$, $\sigma$ vary in $(0, 180^\circ)$. For some values of the lightest neutrino mass,
texture zero $|\langle m\rangle^{}_{\alpha\beta}| =0$ is allowed, 
either in the normal hierarchy (e.g., $|\langle
m\rangle^{}_{ee}| =0$ \cite{Xing03}) or in the inverted hierarchy
(e.g., $|\langle m\rangle^{}_{\tau\tau}| =0$), or in both of them
(e.g., $|\langle m\rangle^{}_{e\mu}| =0$). 

%%%%%%%%%%%%%%%%%%%%%%%%%%%%%%%%%%%%%%%%%%
\section{CP violation in neutrino-antineutrino oscillations}

To eliminate the $|K|^2/E^2$ and
$|\overline{K}|^2/E^2$ factors, we define the CP-violating asymmetry
between $\nu^{}_\alpha \to \overline{\nu}^{}_\beta$ and
$\overline{\nu}^{}_\alpha \to \nu^{}_\beta$ oscillations \cite{Xing13}
\vspace{-0.4cm}
\begin{equation}
{\cal A}^{}_{\alpha\beta} =
\frac{P(\nu^{}_\alpha \to
\overline{\nu}^{}_\beta) - P(\overline{\nu}^{}_\alpha \to
\nu^{}_\beta)}{P(\nu^{}_\alpha \to
\overline{\nu}^{}_\beta) + P(\overline{\nu}^{}_\alpha \to
\nu^{}_\beta)} = \frac{2 \sum^{}_{i<j} m^{}_i m^{}_j {\cal V}^{ij}_{\alpha\beta} \sin
2\phi^{}_{ji}}{|\langle m
\rangle^{}_{\alpha\beta}|^2 - 4\sum^{}_{i<j} m^{}_i m^{}_j
{\cal C}^{ij}_{\alpha\beta} \sin^2\phi^{}_{ji}} \; .
%     (3)
\end{equation}
\vspace{-0.5cm}\\
Only six of the nine CP-violating asymmetries are
independent and nontrivial because of ${\cal A}^{}_{\alpha\beta} = {\cal A}^{}_{\beta\alpha}$. 
The $\nu^{}_\alpha \to \overline{\nu}^{}_\alpha$
oscillation is actually a kind of ``appearance" process and thus it
can accommodate the CP-violating effects. 
Because of the fact $|\Delta m^2_{31}| \simeq |\Delta
m^2_{32}| \simeq 32 \Delta m^2_{21}$, there may exist two
oscillating regions dominated respectively by $\Delta m^2_{21}$ and
$\Delta m^2_{31}$. We classify our analysis into three cases: (1) the normal neutrino mass hierarchy with $m^{}_1 = 0$, (2) the inverted neutrino mass hierarchy with $m^{}_3 = 0$, and (3) the nearly degenerate mass hierarchy with $m^{}_1 \simeq m^{}_2
\simeq m^{}_3$. By fixing the CP phases, we schematically show some results in Figs. (2), (3) and (4), where the inputs of mixing angles and mass-squared differences are the same as above. Note that some of ${\cal A}^{}_{\alpha\beta}$ are remarkably dependent on the CP phases. This has been considered in Ref. \cite{XZ13}, and more detailed discussions can be found over there.

\section{Summary}

In principle, one may determine the Majorana phases of the PMNS matrix in
neutrino-antineutrino oscillations.
Such an experiment might be feasible in the very
distant future, but a systematic study of CP violation
in neutrino-antineutrino oscillations is still useful, so as to
enrich the Majorana neutrino phenomenology. Six
independent $0\nu\beta\beta$-like mass terms $\langle
m\rangle^{}_{\alpha\beta}$ and nine independent Jalskog-like
parameters ${\cal V}^{ij}_{\alpha\beta}$ have been analyzed in
detail, and allowed parameter spaces for $\langle
m\rangle^{}_{\alpha\beta}$ have been presented. We have also
carried out an analysis of the sensitivities of six
CP-violating asymmetries ${\cal A}^{}_{\alpha\beta}$ to the
three phase parameters and the ratio $E/L$ in different mass hierarchies. Our
analytical and numerical results provide a complete description of
the distinct roles of Majorana CP-violating phases in
neutrino-antineutrino oscillations.

\section*{Acknowledgement}
The author is indebted to Z.Z. Xing for his collaboration. This work was supported in part by the National Natural Science Foundation of
China under Grant No. 11135009.

%%%%%%%%%%%%%%%%%%%%%%% Figures %%%%%%%%%%%%%%%%%%%%%%
\begin{figure}[h]
\vspace{-0.6cm}
\begin{minipage}{18.5pc}
\includegraphics[width=19pc, height=21pc]{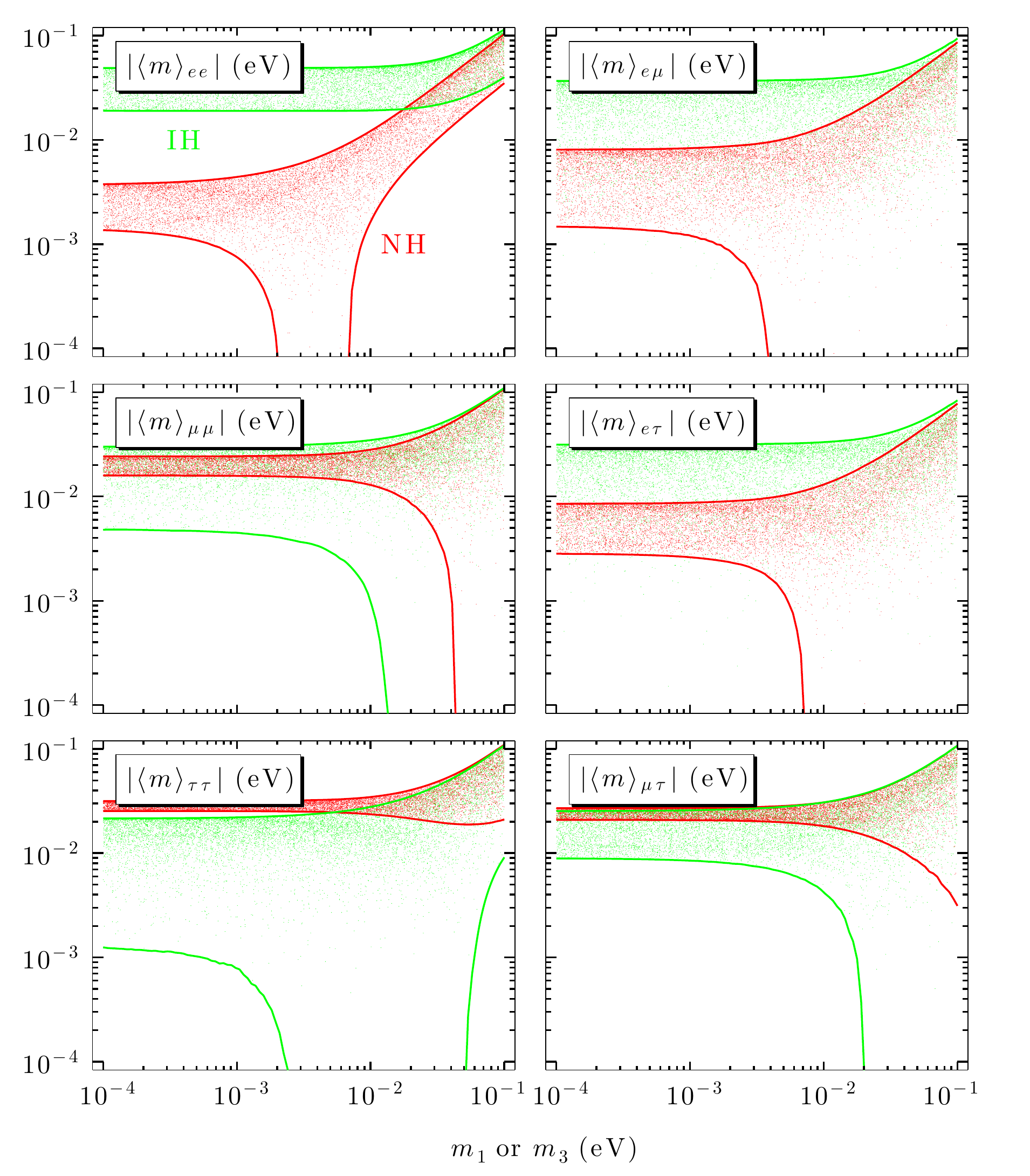}
\vspace{-0.8cm}
\caption{\label{label}The profiles of $|\langle
m\rangle^{}_{\alpha\beta}|$ versus the lightest neutrino mass
$m^{}_1$ (normal hierarchy or NH: red region) or $m^{}_3$ (inverted
hierarchy or IH: green region).}
\end{minipage}
\hspace{0.8pc}%
\begin{minipage}{18.5pc}
\includegraphics[width=19pc, height=21pc]{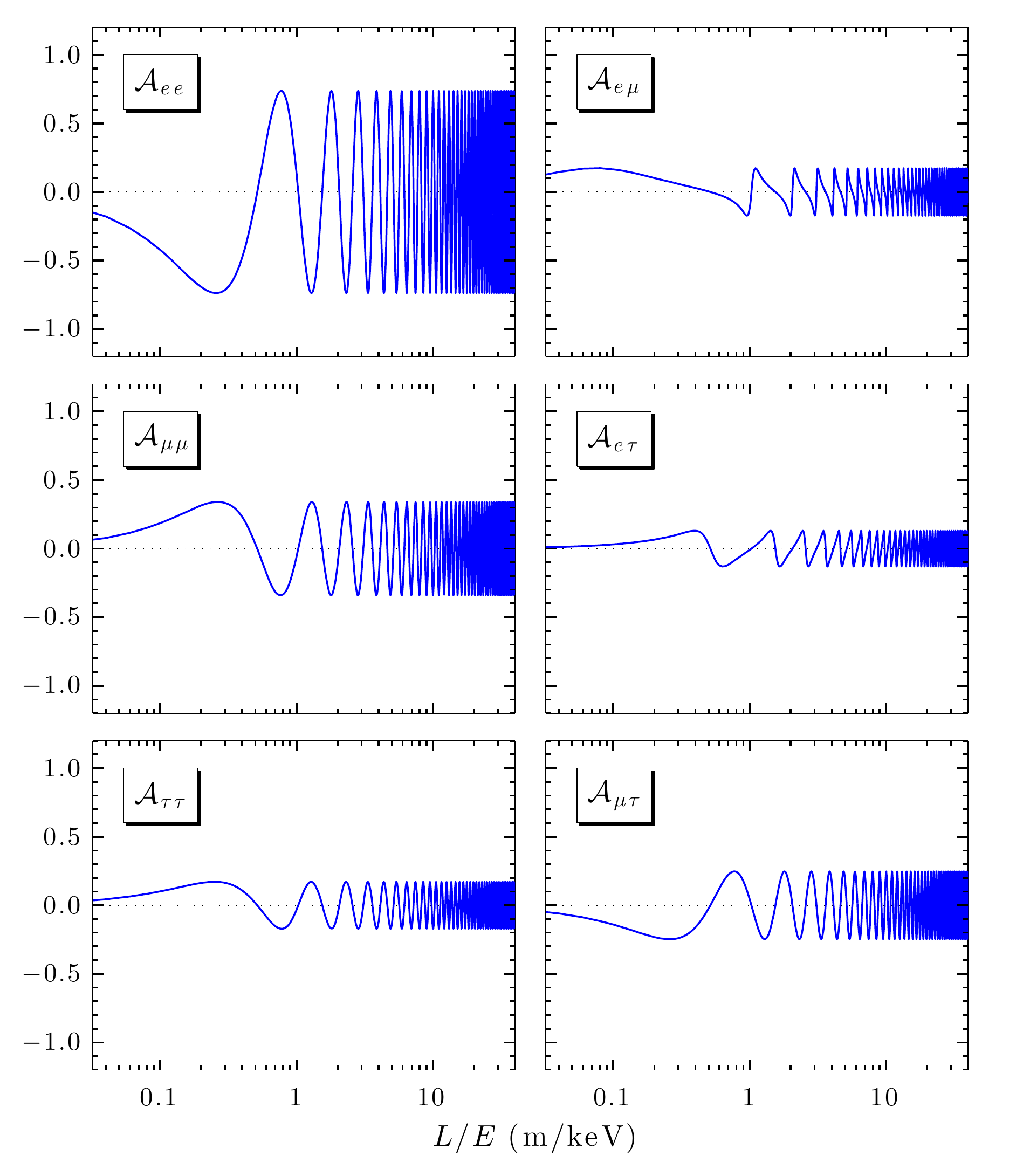}
\vspace{-0.8cm}
\caption{\label{label}The CP-violating asymmetries
$\mathcal{A}^{}_{\alpha\beta}$ versus $L/E$ in the normal neutrino
mass hierarchy with $m^{}_1 =0$, $\delta = 90^\circ$, $\sigma =
45^\circ$, and $\rho$ being arbitrary.}
\end{minipage} 
\vspace{0.3cm}\\
\begin{minipage}{18.5pc}
\includegraphics[width=19pc, height=21pc]{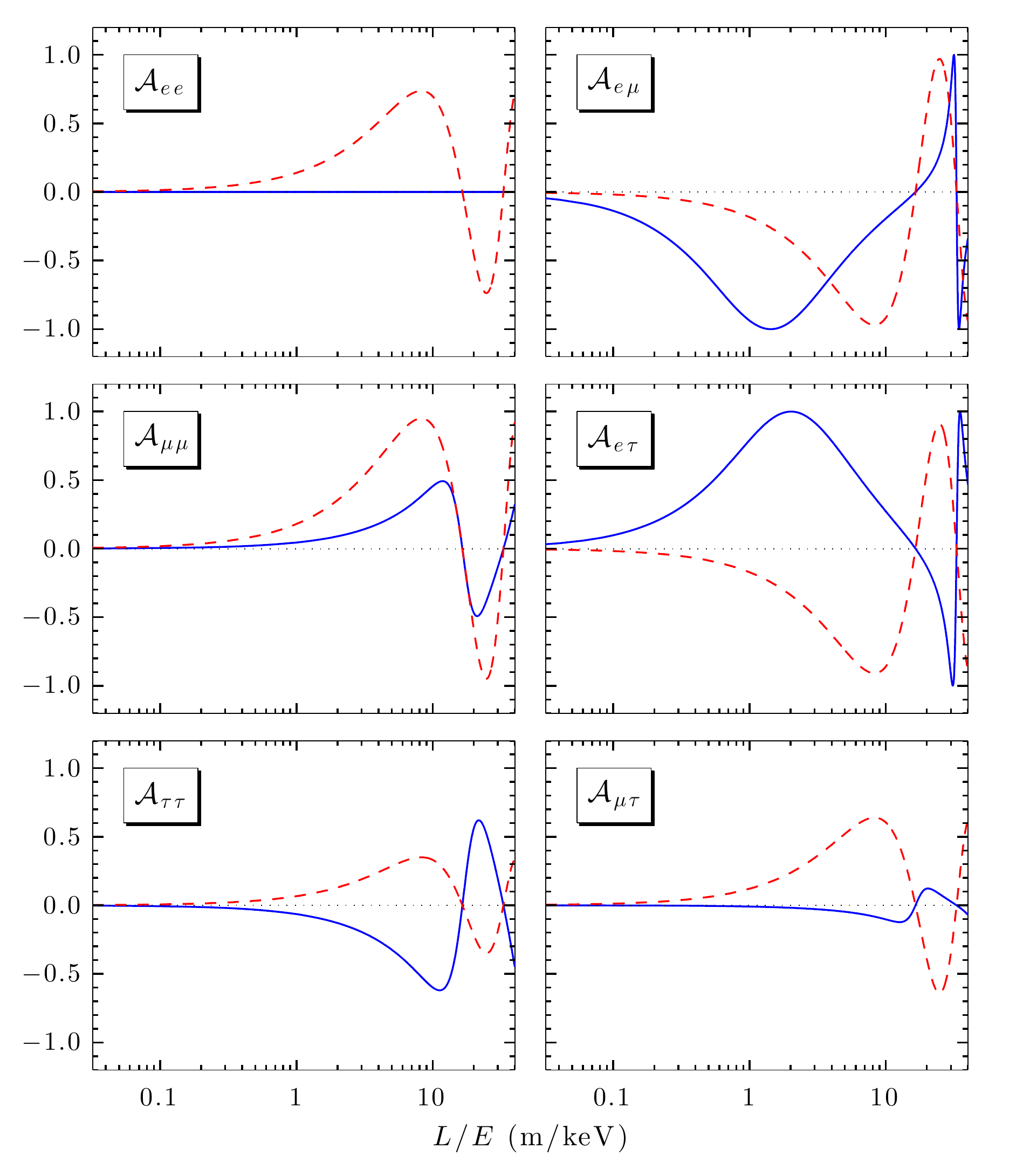}
\vspace{-0.8cm}
\caption{\label{label}The CP-violating asymmetries
$\mathcal{A}^{}_{\alpha\beta}$ versus $L/E$ in the inverted neutrino
mass hierarchy with $m^{}_3 =0$: (a) $\delta = 0^\circ$ and
$\rho-\sigma = 45^\circ$ (red dashed lines); (b) $\delta = 90^\circ$
and $\rho-\sigma = 0^\circ$ (blue solid lines). The absolute values of $\rho$ and $\sigma$ are set to be arbitrary.}
\end{minipage}
\hspace{0.8pc}%
\begin{minipage}{18.5pc}
\includegraphics[width=19pc, height=21pc]{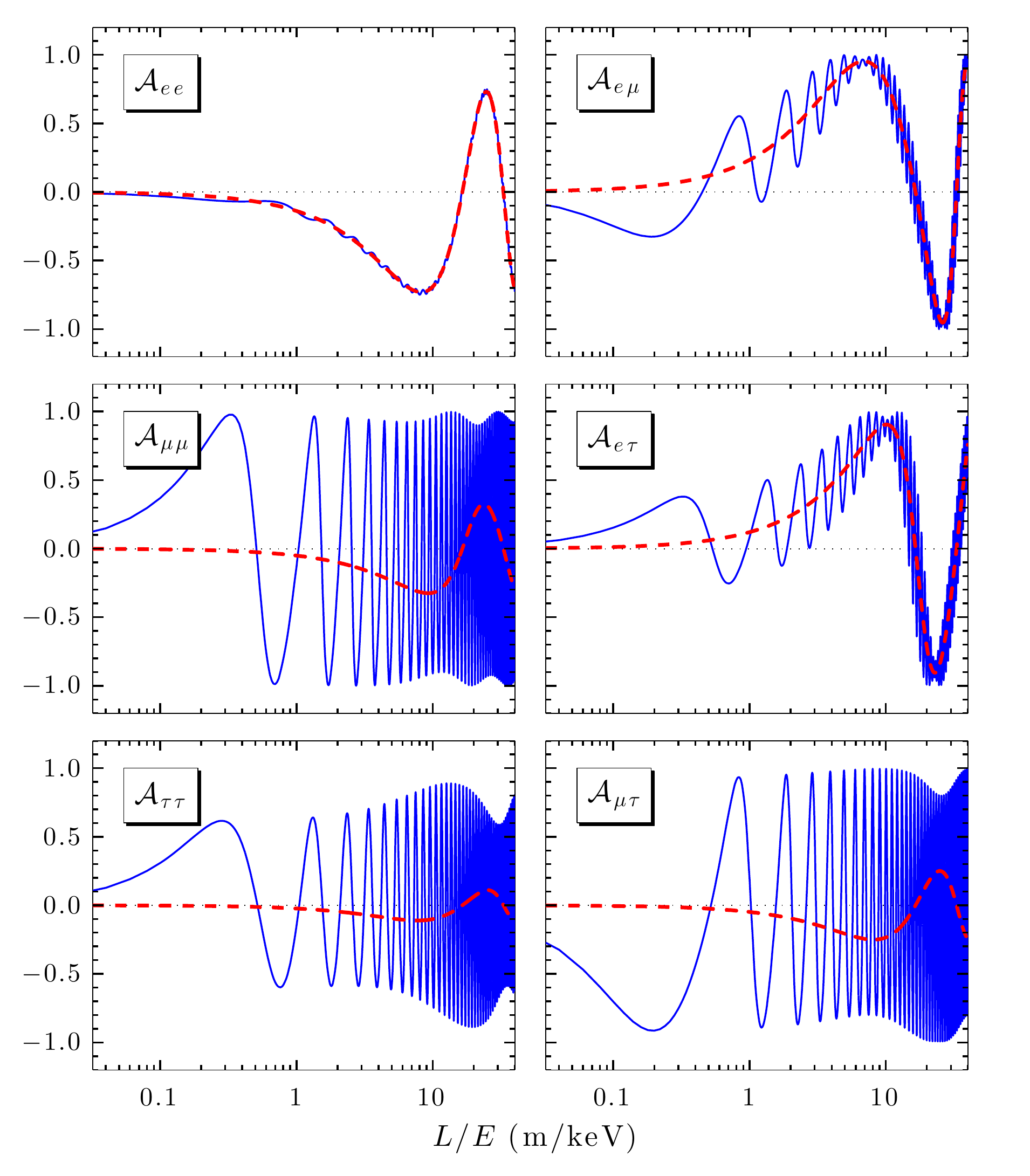}
\vspace{-0.8cm}
\caption{\label{label}The CP-violating asymmetries
$\mathcal{A}^{}_{\alpha\beta}$ (blue solid lines) versus $L/E$ in
the nearly degenerate neutrino mass hierarchy with $m^{}_1 \simeq
m^{}_2 \simeq m^{}_3$, $\rho = 0^\circ$, $\sigma = 45^\circ$ and
$\delta = 90^\circ$, where the red dashed lines stand for the oscillations driven by
$\Delta m^2_{31}$ and $\Delta m^2_{32}$ being averaged out.}
\end{minipage} 
\vspace{-0.5cm}
\end{figure}
%%%%%%%%%%%%%%%%%%%%%%%%%%%%%%%%%%%%%%%%%%


\section*{References}
\begin{thebibliography}{99}

\bibitem{PMNS} Z. Maki, M. Nakagawa, and S. Sakata,
Prog. Theor. Phys. {\bf 28}, 870 (1962);
B. Pontecorvo, Sov. Phys. JETP {\bf 26}, 984 (1968).%

\bibitem{PDG} J. Beringer {\it et al.} (Particle Data Group),
Phys. Rev. D {\bf 86}, 010001 (2012).%

\bibitem{Xing13} Z.Z. Xing, Phys. Rev. D {\bf 87}, 053019 (2013).%

\bibitem{XZ13} Z.Z.~Xing and Y.-L.~Zhou,
  Phys.\ Rev.\ D {\bf 88}, 033002 (2013).%

\bibitem{J} C. Jarlskog, Phys. Rev. Lett. {\bf 55}, 1039 (1985);
D.D. Wu, Phys. Rev. D {\bf 33}, 860 (1986).%

\bibitem{FIT} M.C. Gonzalez-Garcia, M. Maltoni, J. Salvado, and T.
Schwetz, JHEP {\bf 1212}, 123 (2012).%

\bibitem{Han} See, e.g., J. Garayoa and T. Schwetz,
JHEP {\bf 0803}, 009 (2008); M. Kadastik, M. Raidal, and L. Rebane,
Phys. Rev. D {\bf 77}, 115023 (2008); A.G. Akeroyd, M. Aoki, and H.
Sugiyama, Phys. Rev. D {\bf 77}, 075010; Z.Z. Xing, Phys. Rev. D
{\bf 78}, 011301 (2008); P. Fileviez Perez, T. Han, G.Y. Huang, T.
Li, and K. Wang, Phys. Rev. D {\bf 78}, 015018 (2008); F. del Aguila
and J.A. Aguilar-Saavedra, Nucl. Phys. B {\bf 813}, 22 (2009).%

\bibitem{SS2} W. Konetschny and W. Kummer, Phys. Lett. B {\bf 70},
433 (1977); M. Magg and C. Wetterich, Phys. Lett. B {\bf 94}, 61
(1980); J. Schechter and J.W.F. Valle, Phys. Rev. D {\bf 22}, 2227
(1980); T.P. Cheng and L.F. Li, Phys. Rev. D {\bf 22}, 2860 (1980);
G. Lazarides, Q. Shafi, and C. Wetterich, Nucl. Phys. B {\bf 181},
287 (1981); R.N. Mohapatra and G. Senjanovic, Phys. Rev. D {\bf 23},
165 (1981).%

\bibitem{Valle} J. Bahcall and H. Primakoff, Phys. Rev. D {\bf 18},
3463 (1978); L.N. Chang and N.P. Chang, Phys. Rev. Lett. {\bf 45},
1540 (1980); J. Schechter and J.W.F. Valle, Phys. Rev. D {\bf 23},
1666 (1981); L.F. Li and F. Wilczek, Phys. Rev. D {\bf 25}, 143
(1982); J. Bernabeu and P. Pascual, Nucl. Phys. B {\bf 228}, 21
(1983); J.D. Vergados, Phys. Rept. {\bf 133}, 1 (1986); P. Langacker
and J. Wang, Phys. Rev. D {\bf 58}, 093004 (1998); A. de Gouvea, B.
Kayser, and R.N. Mohapatra, Phys. Rev. D {\bf 67}, 053004 (2003); D.
Delepine, V.G. Macias, S. Khalil, and G.L. Castro, Phys. Lett. B
{\bf 693}, 438 (2010).%

\bibitem{Xing03} Z.Z. Xing, Phys. Rev. D {\bf 68}, 053002 (2003);
Y.F. Li and S.S. Liu, Phys. Lett. B {\bf 706}, 406 (2012).%



\end{thebibliography}
\end{document}